\documentclass[aps,prb,twocolumn,nofootinbib]{revtex4-2}

\usepackage{graphicx}
\usepackage{dcolumn}
\usepackage{bm}
\usepackage{textcomp}
\usepackage{color}
\usepackage{amsmath}
\usepackage{amssymb}
\usepackage{xcolor}
\usepackage{hyperref}
\usepackage{easyReview}
\usepackage{mathdots}
\usepackage{float}
\usepackage{lineno}
\usepackage{array}
\usepackage{xcolor,colortbl}
\usepackage[normalem]{ulem}
\definecolor{LightBlue}{rgb}{0.8,0.8,0.8}
\allowdisplaybreaks
\usepackage{amsmath}
\usepackage{lineno}

\newcommand{\eph}{{\it e}-ph}

\begin{document}
\title{Self-learning QMC: application to the classical Holstein-Spin-Fermion model}
\author{Shaozhi Li}
\email{lishaozhi@mail.neu.edu.cn}
\affiliation{Department of Physics, Northeastern University, Shenyang 110819, China}

\begin{abstract}
To evaluate the effectiveness of machine learning in systems with competing interactions, we developed a self-learning quantum Monte Carlo (SLQMC) method to simulate the phase transition in the classical Holstein–spin-fermion model. In SLQMC, machine learning techniques are employed to approximate the free energy, thereby bypassing the need for exact diagonalization and significantly reducing computational cost. We assess the performance of SLQMC using both linear regression and neural network models. Our results show that both models are capable of capturing the phase transition from the antiferromagnetic state to the charge-density-wave state. However, the sampling efficiency decreases near the AFM-CDW phase transition, which is attributed to the increased mean-squared-error of the machine learning model. Additionally, the sampling efficiency decreases with increasing lattice size. This suppression is due to the increased root-mean-squared-error as the machine learning model is applied to a large lattice  and the finite-size effect, wherein the energy gap between the ground state and low-energy excited states decreases as the lattice grows. Our findings highlight the necessity of highly accurate machine learning models to simulate theoretical models with complex, competing microscopic interactions on a large lattice.
\end{abstract}

\maketitle

\section{Introduction}
Machine learning techniques are playing a pivotal role in accelerating research in material science and condensed matter physics~\cite{Schmidt2019recent,Faber2016Machine,Schmidt2017Predicting,Schutt2014how,Kim2018machine,Wei2018predicting,Ye2018Deep,Nagai2020self}. For instance, AlphaFold has speeded up to predict protein structures, increasing the structural coverage from 48\% to 76\% of all human protein residues~\cite{Jumper2021highly}. In addition, the deep learning method has been applied to analyzing inelastic neutron scattering spectroscopies of spin ice $\text{Dy}_2\text{Ti}_2\text{O}_7$~\cite{Samarakoon2020Complex}. Furthermore, the machine learning method has been adopted to predict the superconducting temperature of quantum materials~\cite{Stanev2018machine,xie2022machine}. These applications exemplify how machine learning is reshaping research paradigms, providing new tools to tackle longstanding challenges in physics and material science.

In the field of condensed matter physics, both supervised and unsupervised machine learning techniques have been widely adopted to solve a variety of physics models with many-body interactions. Unsupervised learning approaches, in particular, have been used to compress quantum states, thereby enhancing computational efficiency~\cite{Nagy2019variational, Vicentini2019Variational, Yang2020deep}. For example, neural network quantum states based on restricted Boltzmann machine (RBM) architectures have been successfully employed to represent the ground states of the transverse-field Ising model and the antiferromagnetic Heisenberg model~\cite{Giuseppe2017Science}. Moreover, a variety of neural network architectures have been developed to represent the ground states of both fermionic~\cite{Nomura2017Restricted, Javier2022Fermionic, nomura2024quantum, Nomura2021helping} and bosonic~\cite{Saitp2017Solving, Saitp2017Solving, Saito2018Machine, Zhu2023HubbardNet} Hubbard models. Furthermore, quantum algorithms based on RBM wave function ans\"atze have been developed to simulate nonequilibrium dynamics of quantum many-body systems~\cite{Lee2021Neural}. In contrast, supervised machine learning techniques have been applied to accelerate quantum Monte Carlo and molecular dynamics simulations. For example, machine learning approaches have been integrated into determinant quantum Monte Carlo (QMC) methods to improve the efficiency of simulating large-scale Holstein models~\cite{Shaozhi2020Accelerating, Chen2018Symmetry, Wynen2021machine}. Supervised neural networks have been used in molecular dynamics to predict nonequilibrium electron forces~\cite{Zhang2023Machine}.

Beyond idealized physics models, machine learning techniques have been integrated into density functional theory (DFT) and molecular dynamics to accelerate simulations of real quantum materials~\cite{Li2007improving, chaofang2020understanding, Hong2020training, Wu2020modeling, Stricker2020machine, Wen2019development}. For instance, NeuralXC was developed to predict exchange-correlation functions in DFT~\cite{Dick2019Machine}. Notably, the NeuralXC outperforms conventional approaches in describing bond breaking of water and shows excellent agreement with experimental results. In molecular dynamics, neural network potentials have been introduced to predict interatomic potentials~\cite{Fiedler2022Deep,Gartner2020signature}. In both DFT and molecular dynamics, machine learning models are typically trained to predict the total energy based on atomic configurations. However, in many earlier applications of machine learning to physics models~\cite{junwei2017self,Evan2021Data}, the lattice degree of freedom are often neglected. While some studies have incorporated the electron-phonon ({\eph}) interaction, they omitted electron-electron interactions~\cite{Shaozhi2020Accelerating, Chen2018Symmetry} .  As a result, the application of machine learning to physics models with both {\eph} and electron-electron interactions remains rare.

Incorporating both the lattice and electron degree of freedoms into physics models can lead to a rich variety of phenomena. For example, the interplay between electron-electron and {\eph} interactions can drive phase transitions between antiferromagnetic (AFM) and charge-density-wave (CDW) states. These competing interactions can also give rise to competition between s-wave and d-wave superconducting orders. Motivated by these phenomena, it is natural to ask whether machine learning approaches can effectively capture phase transitions arising from such competing interactions. However, this question remains largely unexplored. For {\it ab initio} approaches, it is challenging to tune interaction strengths, making it difficult to systematically evaluate the performance of machine learning methods in capturing phase transitions induced by competing interactions. In contrast, model studies offer more control. For instance, a machine learning-assisted Langevin dynamics approach was recently introduced to simulate CDW dynamics of the Hubbard-Holstein model, demonstrating the potential of machine learning for modeling the interplay of multiple microscopic interactions~\cite{Yang2024Enhanced}. Nevertheless, Ref.~\cite{Yang2024Enhanced} focuses exclusively on the CDW regime with strong {\eph} coupling, leaving the performance near the phase transition region uncertain.

The {\it objective} of this work is to evaluate the effectivenesee of machine learning approaches in handling competing interactions. To this end, we develop a self-learning quantum Monte Carlo (SLQMC) method to study phase transitions in the classical Holstein-spin-fermion model, in which the interplay between the spin-spin and {\eph} interactions drives a phase transition between the AFM and CDW states. This model is typically studied using the conventional QMC method, which is a computationally expensive due to the need for exact diagonalization (ED) to evaluate the free energy at each Monte Carlo update step. To overcome this limitation, SLQMC employees a machine learning model to predict the free energy, thereby bypassing exact diagonalization and significantly reducing the computational cost. We explore two machine learning models: a linear regression (LR) model and a neural network (NN) model. Our results show that SLQMC performs well in both the strongly AFM-fluctuating regime, where the {\eph} interaction is weak, and the CDW regime, where the {\eph} interaction is strong. However, the sampling efficiency decreases near the AFM-CDW phase transition, which is attributed to the increased mean-squared-error (MSE) of the machine learning model. Additionally, we observe that the sampling efficiency decreases with increasing lattice size. This suppression is due to the increased root-mean-squared-error (RMSE) as the machine learning model is applied to a large lattice and the finite-size effect, wherein the energy gap between the ground state and low-energy excited states decreases as the lattice grows.

\begin{figure*}[ht]
\centering
\includegraphics[width=0.7\textwidth]{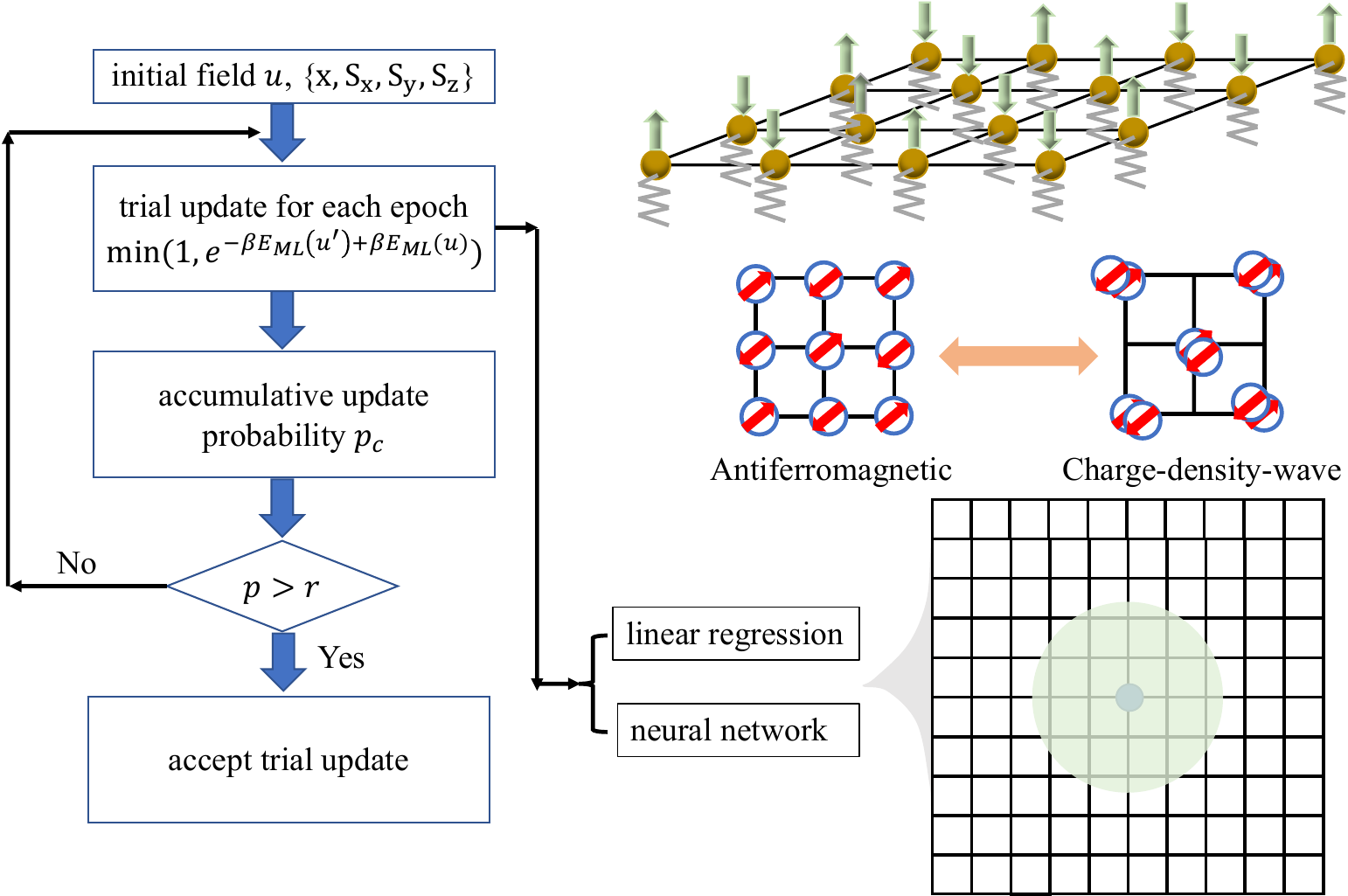}
\caption{The left panel sketches self-learning QMC. The upper right panel shows a fermion model with additional spin and lattice degrees of freedom.
}
\label{Fig:fig1}
\end{figure*}

This paper is organized as follows. In Sec.\ref{sec:II}, we introduce the classical Holstein-spin-fermion model and describe the SLQMC method. Sec.\ref{sec:III} begins with a discussion of a linear regression model and its corresponding training results, followed by the application of SLQMC. We then introduce a neural network model, which has a smaller loss function compared to the linear regression model. Additionally, we present size-dependent SLQMC results at low temperatures near the phase transition point using the neural network model. We explained the origin of the sampling inefficiency on a large lattice size. Furthermore, we present the success of SLQMC across low to high temperatures. Finally, in Sec.~\ref{sec:IV}, we provide the conclusions of our work.

\section{Model and Methodology}\label{sec:II}
\subsection{Classical Holstein-Spin Fermion model}
To evaluate the effectiveness of the machine learning-assisted approach for a complex system, we consider a theoretical model that includes two interactions. A prototypical microscopic model for such systems is the Hubbard-Holstein model, which captures the interplay between {\eph} and electron-electron interactions. The corresponding Hamiltonian is defined as
\begin{eqnarray}
H=H_0+H_\mathrm{\eph}+H_\mathrm{e-e}+H_\mathrm{lattice},
\end{eqnarray}
where $H_0$ is the tight-binding model, $H_\mathrm{\eph}$ represents the {\eph} interaction, $H_\mathrm{e-e}$ accounts for the electron-electron interaction, and $H_\mathrm{lattice}$ describes the lattice Hamiltonian. For the Hubbard-Holstein model, these interactions are given by the following expressions:
\begin{eqnarray}
H_0&=&-t\sum_{\langle i, j \rangle,\sigma} c_{i,\sigma}^{\dagger} c_{j,\sigma}^{\phantom\dagger} - \mu\sum_{i,\sigma} \hat{n}_{i,\sigma}\\
H_\mathrm{\eph}&=&g\sum_{i,\sigma} \hat{x}_i (\hat{n}_{i,\sigma}-\frac{1}{2})\\
H_\mathrm{e-e}&=&U\sum_{i} \hat{n}_{i,\uparrow} \hat{n}_{i,\downarrow}\\
H_\mathrm{lattice}&=&\sum_i \left[\frac{1}{2M}\left(\frac{d\hat{x}_i}{dt}\right)^2 + \frac{K^2}{2} \hat{x}_i^2\right].
\end{eqnarray}
Here, $t$ is the nearest-neighbor (NN) hopping integral, $\mu$ is the chemical potential controlling the electron density, $g$ denotes the {\eph} interaction strength, and $U$ is the onsite Coulomb interaction. The parameters $M$ and $K$ correspond to the mass and the elastic constant of the lattice, respectively. Throughout this paper, we set $t=1$ and $K=1$, while the chemical potential is adjusted to maintain an electron density of 1.

Many previous studies have extensively employed the QMC method to study this Hamiltonian~\cite{Clay2005Intermediate, nowadnick2012competition,Pradhan2015Holstein, shaozhi2018phase,Hebert201one,Huang2022determinantal}. The presence of both the {\eph} and electron-electron interactions induces the notorious sign problem, which persists even at half-filling. To get rid of this issue, we employ a mean-field approximation and transform the Hubbard-Holstein model into the classical Holstein-spin-fermion (HSF) model. The resulting Hamiltonian is expressed as
\begin{eqnarray}
H=H_0+H_\mathrm{\eph}+H_\mathrm{sf}+H_\mathrm{lattice},
\end{eqnarray}
where $H_\mathrm{sf}$ denotes the spin-fermion interaction, defined as
\begin{eqnarray}
H_\mathrm{sf}=J\sum_{i} \bf{S}_i \cdot \hat{\bf{s}}_i.
\end{eqnarray}
Here, $\bf{S}_i$ represents a local classical spin vector at site $i$, whose amplitude is 1. $\hat{\bf{s}}_i$ denotes a quantum spin operator at site $i$. The coupling strength $J$ is set to $J=2$. In the adiabatic limit ($M\rightarrow \infty$), the lattice Hamiltonian simplifies to 
\begin{eqnarray}
H_\mathrm{lattice}=\sum_i \frac{K^2}{2} x_i^2,
\end{eqnarray}
where $x_i$ is a classical value rather than a quantum operator. 

\subsection{Self-learning QMC}
Next, we discuss the self-learning QMC method, initially proposed in Ref.~\cite{Liu2017self} to simulate a classical spin model. Here, we extend this method to study the classical HSF model. 

To simulate the HSF model using QMC, we need to include four auxiliary fields: $x_i, S_{i,x}, S_{i,y}$, and $S_{i,z}$. For simplicity, we label the collection of these fields as $u$. In the QMC simulation, each epoch includes updates of all these four auxiliary fields. The acceptance probability of changing a field from $u$ to $u^\prime$ is given by $p=\mathrm{min}(1,p_u)$, where
\begin{eqnarray}
p_u= e^{-\beta F(u^\prime)+\beta F(u)}.
\end{eqnarray}
Here, $\beta$ is the inverse temperature. $F(u)$ denotes the free energy of the system, which is expressed as
\begin{eqnarray}\label{eq: freeE}
F(u)&=&-\sum_{i}\mathrm{ln}(1+e^{-\beta E_i})/\beta+E_\mathrm{lattice},\\
E_\mathrm{lattice}&=&\sum_i \frac{K^2}{2}x_i^2 - g\sum_i x_i,
\end{eqnarray}
where $E_i$ is the eigen-energy of the system with auxiliary field $u$. To obtain these eigen-energies, we performed exact diagonalization (ED) using Lapack zgeev routine. However, this ED approach is inefficient because its computational complexity is $\mathcal{O}(N^3)$, where $N$ is the size of the system. Therefore, for each epoch of QMC, the computational complexity is $\mathcal{O}(N^4)$.

To improve the efficiency of the QMC method, we adopt machine learning methods to evaluate the free energy. The flowchart of the self-learning QMC method is sketched in Fig.~\ref{Fig:fig1}. Considering an initial field $u$, we first perform a trial update with the probability
\begin{eqnarray}
p_\mathrm{trial}=\mathrm{min}(1, e^{-\beta F_{ML}(u^\prime) + \beta F_{ML}(u)}),
\end{eqnarray}
where $F_\mathrm{ML}(u)$ is the estimated free energy from machine learning methods using input $u$. In each epoch, this trial update should be performed through all auxiliary fields at each site. After each epoch, we need to perform a cumulative update with the probability
\begin{eqnarray}
p_c=\mathrm{min}(1, e^{-\beta (F(u^\prime)-F(u))} e^{\beta \sum_i \Delta F_i}),
\end{eqnarray}
where $F(u)$ ($F(u^\prime)$) denotes the free energy of the system at the start (end) of each epoch. The symbol $\Delta F_i$ denotes the predicted changed free energy for the accepted trial updates. In SLQMC, the computational complexity of each epoch is $\mathcal{O}(N^3)$, which is smaller than the complexity of the standard QMC, $\mathcal{O}(N^4)$.

In QMC simulations, the update of auxiliary fields could be stuck due to the presence of the local minimum energy. This issue becomes severe at low temperatures. To address this issue, we adopt the annealing method ~\cite{Karimi2017effective, Chen2022optimizing} so that the system is gradually cooled from a high temperature ($T=t$) to a low temperature ($T=0.1t$). In our simulations, the temperature decay rate is 0.9, and we set 2000 epochs at each temperature. This temperature annealing method is also adopted in SLQMC simulations.

\section{Results}\label{sec:III}
The success of SLQMC depends on the high accuracy of the machine learning model. In our work, we adopt linear regression and neural network models to predict free energy.
\subsection{Linear regression model}
We first discuss the linear regression model. We assume that the free energy of the HSF model can be approximated using the following equation,
\begin{eqnarray}\label{eq: LRfunc}
F_\mathrm{LR}&=& E_\mathrm{s-s}+E_\mathrm{l-l}+E_\mathrm{s-l}+W(x)+c_0\\
E_\mathrm{s-s}&=& \sum_{i,{\bf d}} J_{\bf d} {\bf S}_{{\bf r}_i} {\bf S}_{{\bf r}_i+{\bf d}} \\
E_\mathrm{l-l}&=&\sum_{i,{\bf d}} K_{\bf d} x_{{\bf r}_i} x_{{\bf r}_i+{\bf d}} \\
E_\mathrm{s-l}&=&\sum_{i,{\bf d}} h_{\bf d} {\bf S}_{{\bf r}_ i} {\bf S}_{{\bf r}_ i+{\bf d}}  x_{{\bf r}_ i} x_{{\bf r}_ i+{\bf d}}\\ 
W(x)&=&\sum_{i} c_1 \left [ \frac{x_{{\bf r}_i}^4}{4} -\frac{ \overline{x^2} x_{{\bf r}_i}^2}{2}   \right] + c_2 x_{{\bf r}_i}^2  + c_3 |x_{{\bf r}_i}|. 
\end{eqnarray}
The first term in Eq.~(\ref{eq: LRfunc}) represents the effective spin-spin interaction, the second term denotes the effective lattice-lattice interaction, and the third term represents the effective spin-lattice interaction. The interaction strengths $J_{\bf d}$, $K_{\bf d}$, and $h_{\bf d}$ are restricted to follow the $C_4$ symmetry on the square lattice. $W(x)$ is an effective onsite elastic energy, which is adopted to constrain the range of motion of the atoms, as explained in Ref.~\cite{Shaozhi2020Accelerating}.The first term in $W(x)$ reflects a double well lattice potential characteristic of a CDW state. The second term captures the harmonic oscillator-like lattice potential, which becomes important in an AFM state. The third term accounts for a linear dependence of the lattice potentials on the displacement, which may arise from the e-ph interaction. The machine learning model automatically determines the weights ($c_1$, $c_2$, and $c_3$) of these three contributions. $\overline{x^2}$ is the average value of $x^2$, evaluated from training samples. In the linear regression model, the effective spin-spin, lattice-lattice, and spin-lattice interactions are included up to the $7$-th NN, that is ${\bf d}=\pm3\hat{x}\pm\hat{y}$ and $\pm\hat{x}\pm3\hat{y}$. We note that including interactions with a longer distance does not improve the training result. Therefore, our linear regression model has 24 terms except for the bias term.

\begin{figure}[t]
\centering
\includegraphics[width=0.99\columnwidth]{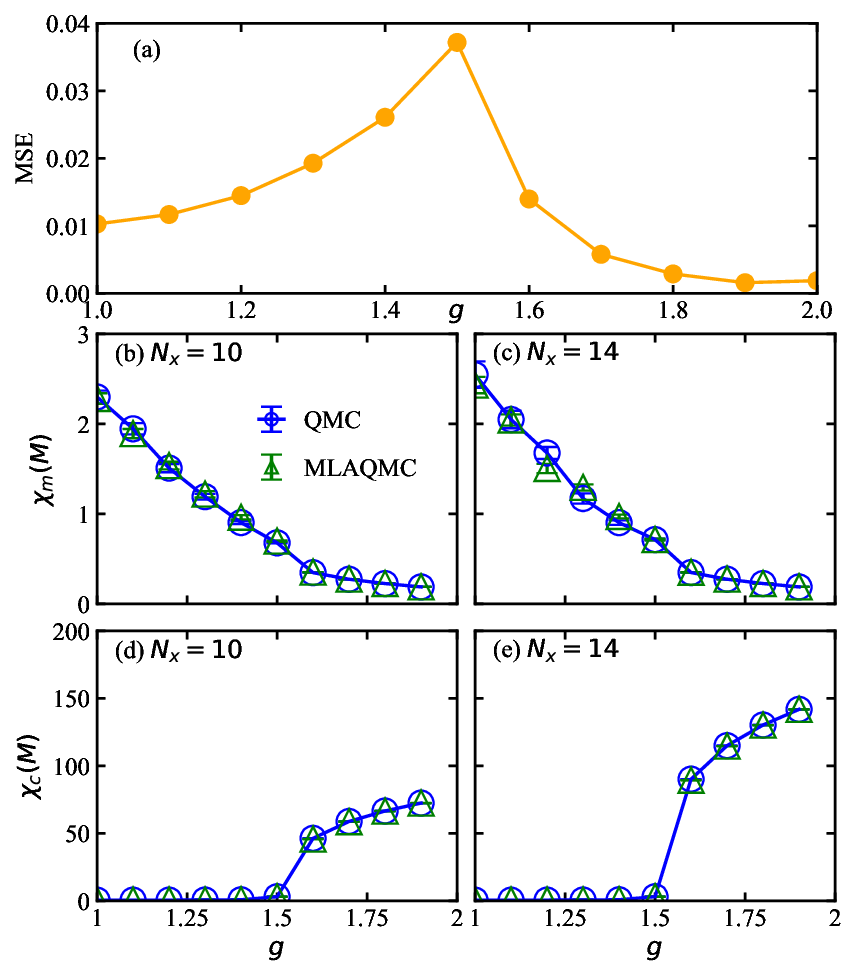}
\caption{ (a) The mean-squared-error (MSE) loss function for a trained linear regression model as a function of the electron-phonon coupling strength $g$. (b)-(c) The magnetic susceptibility $\chi_m$ at the momentum point $M$ as a function of $g$. (d)-(e) The charge susceptibility $\chi_c$ at the momentum point $M$ as a function of $g$. The temperature is set as $\beta=10$. 
}
\label{Fig:fig2}
\end{figure}

We trained this linear regression model to predict the free energy of the HSF model with $J=2$ for various {\eph} interaction strengths at $\beta=10$. The training dataset consists of 168,000 samples, while the testing dataset includes 72,000 samples. All samples are generated from QMC simulations of the HSF model on a $10\times10$ lattice. Our numerical results indicate that a phase transition from an AFM state to a CDW state occurs around $g=1.5$ (see details in App.~\ref{sec:app1}). At $\beta=10$, the system exhibits AFM fluctuations for $g<1.5$ and CDW fluctuations for $1.5\le g < 1.8$. Upon lowering the temperature, these fluctuating states develop into long-range ordered phases (see details in Apps.~\ref{sec:app4} and~\ref{sec:app5}). For simplicity, we will hereafter refer to the regime $g<1.5$ as the AFM region and $g>1.5$ as the CDW region. Figure~\ref{Fig:fig2}(a) shows the MSE loss function of the trained linear regression model evaluated on the testing dataset for different values of $g$. These MSE loss functions are below 0.04 across various {\eph} interaction strengths; therefore, the RMSE is smaller than 0.2, which is significantly smaller than the ground state free energy of the system (approximately $-200$), indicating that our model can accurately capture the ground state free energy of the HSF model. Additionally, the MSE loss function gradually increases as $g$ varies from 1 to 1.5 and then rapidly decreases as $g$ increases further. The reduced accuracy of the linear regression model near the phase transition point ($g = 1.5$) is attributed to the complex fluctuations of spin and lattice degrees of freedom, which require a more sophisticated machine learning model to capture these dynamics fully.

\begin{figure}[t]
\centering
\includegraphics[width=0.99\columnwidth]{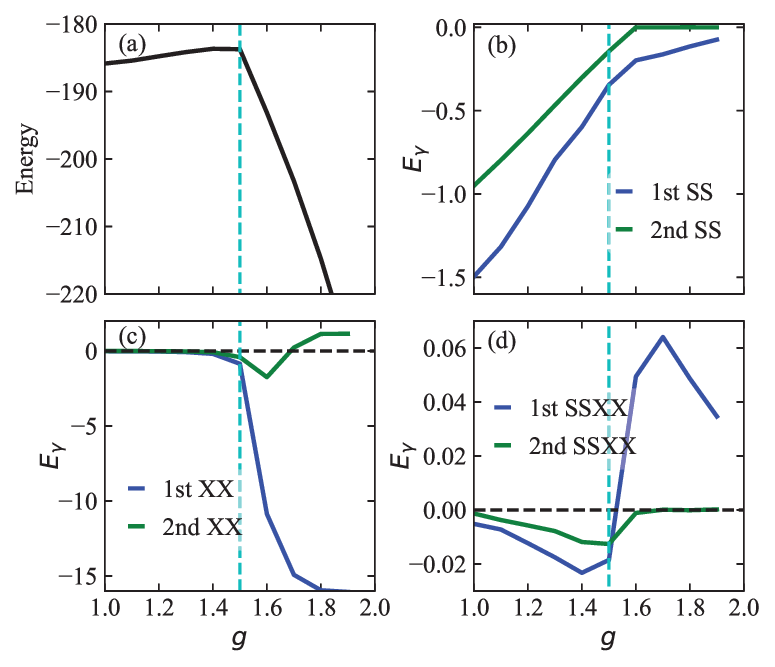}
\caption{ (a) The energy of the system as a function of the electron-phonon interaction $g$. (b) The energy of spin-spin interactions from the linear regression model. (c) The energy of lattice-lattice interactions from the linear regression model. (d) The energy of spin-lattice interactions from the linear regression model. The temperature is set as $\beta=10$.
}
\label{Fig:fig3}
\end{figure}

To further examine this linear regression model, we integrate it into the QMC method to predict the free energy. While the linear regression model is trained to predict the free energy of a $10\times 10$ lattice, we estimate the free energy of a lattice with a size of $N$ using $\frac{N}{100}F_\mathrm{LR}$. Figures~\ref{Fig:fig2}(b)-~\ref{Fig:fig2}(e) plot the magnetic susceptibility $\chi_m(M)$ and charge susceptibility $\chi_c(M)$ at the momentum point $M$ for the $10\times 10$ and $14\times 14$ lattices. 
Here, the magnetic susceptibility $\chi_m({\bf q})$ and the charge susceptibility $\chi_c({\bf q})$ are defined as
\begin{eqnarray}
\chi_m({\bf q})=\frac{1}{N} \sum_{i,{\bf d}} \langle \hat{\bf s}_{{\bf r}_i} \hat{{\bf s}}_{{\bf r}_i+{\bf d}} \rangle e^{i{\bf q}\cdot {\bf d}}
\end{eqnarray}
and
\begin{eqnarray}
\chi_c({\bf q})=\frac{1}{N} \sum_{i,{\bf d}} \langle \hat{n}_{{\bf r}_i} \hat{n}_{{\bf r}_i+{\bf d}} \rangle, e^{i{\bf q}\cdot {\bf d}},
\end{eqnarray}
respectively. The circle and triangular symbols denote the results of the standard QMC and SLQMC with the linear regression model, respectively. As shown in Figs~\ref{Fig:fig2}(b)-~\ref{Fig:fig2}(e), the results of SLQMC are consistent with the results of QMC, implying that our SLQMC can be applied to study the phase transition induced by the competition between two interactions.

Next, we analyze the energy contributions of different interactions across the phase transition based on the linear regression model. Figure~\ref{Fig:fig3}(a) illustrates the variation of the total free energy as the system transitions between AFM and CDW phases for a $10\times 10$ lattice. In the AFM region ($g<1.5$), the total energy is slightly increased by the {\eph} interaction, whereas in the CDW region, the {\eph} interaction rapidly reduces the total energy. The increase in energy in the AFM region is due to spin fluctuations, which excite the AFM ground state to a higher-energy magnetic state. In contrast, the sharp decrease in energy in the CDW region is attributed to the {\eph} coupling term $gx\hat{n}$. In the CDW state, atoms are displaced from their original positions by a distance of $\pm a$, where $a$ is a positive value. The electron density on atoms displaced by $a$ becomes zero, while the electron density on atoms displaced by $-a$ becomes two. As a result, the {\eph} interaction lowers the energy of the system by $Nga$.

Figure~\ref{Fig:fig3}(b) shows the effective spin-spin interaction energies for the first and second nearest neighbors (NN). Both interaction energies decrease rapidly in the AFM region, indicating that the {\eph} interaction suppresses the magnetic state. In the CDW region, the second NN spin-spin interaction energy drops to zero, while the first NN spin-spin interaction energy remains nonzero. This result suggests that spin-spin fluctuations do not vanish in the CDW state but are confined to nearest neighbors only. Similarly, Fig.~\ref{Fig:fig3}(c) 
shows the effective lattice-lattice interaction energies for the first and second NN. In the AFM region, the first NN lattice-lattice interaction energy is negligible but decreases rapidly in the CDW region due to changes in the displacement $a$. In contrast, the second NN lattice-lattice interaction energy is much smaller in the CDW region and fluctuates around zero. These fluctuations arise from lattice vibrations induced by the temperature effect. For a CDW state without lattice fluctuations, the second NN lattice correlation is expected to be zero. Figure~\ref{Fig:fig3}(d) shows the effective spin-lattice interaction energies for the first and second nearest neighbors (NN). The results indicate that spin-lattice interactions are present across the phase transition, but their magnitudes are much smaller compared to the effective spin-spin and lattice-lattice interaction energies. Interestingly, the first NN spin-lattice interaction energy is negative in the AFM region and becomes positive in the CDW region, suggesting that the spin-lattice interaction suppresses the formation of the CDW state. In contrast, the second NN spin-lattice interaction energy vanishes in the CDW region, which can be attributed to the absence of second NN spin-spin correlations.

\begin{figure}[t]
\centering
\includegraphics[width=0.99\columnwidth]{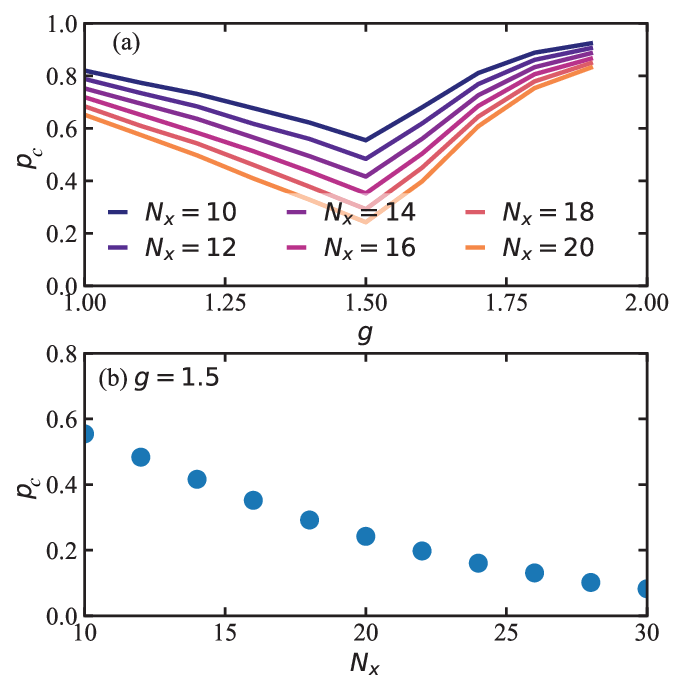}
\caption{ (a) The cumulative update probability $p_c$ as a function of the electron-phonon interaction $g$ for different lattice sizes. (b) The cumulative update probability $p_c$ as a function of the lattice size at $g=1.5$. These results are obtained from the self-learning QMC method using the linear regression model. The temperature is set as $\beta=10$.
}
\label{Fig:fig4}
\end{figure}

Next, we analyze the efficiency of the sampling method based on the linear regression model. If the samples generated from the linear regression model are accurate, the cumulative update probability $p_c$ should approach 1; otherwise, $p_c$ will be close to zero. Figure~\ref{Fig:fig4}(a) shows the variation of $p_c$ across the phase transition for different lattice sizes $N = N_x^2$. For $N_x = 10$, $p_c$ is suppressed by the {\eph} interaction in the  AFM region and enhanced in the charge-density-wave (CDW) region. As the lattice size increases, $p_c$ is suppressed for all values of $g$. However, the suppression is relatively minor at $g = 1.0$ and $g = 1.9$, while the most significant suppression occurs near the phase transition point ($g = 1.5$). To further elucidate this suppression, we plot the size dependence of $p_c$ at $g = 1.5$ in Fig.~\ref{Fig:fig4}(b). It is observed that $p_c$ decreases smoothly from 0.6 to 0.1 as $N_x$ increases from 10 to 30. These results indicate that while the linear regression model can generate good samples for the Holstein and spin-fermion models, it lacks efficiency for simulating a large system near the phase transition point.

\begin{figure}[t]
\centering
\includegraphics[width=0.99\columnwidth]{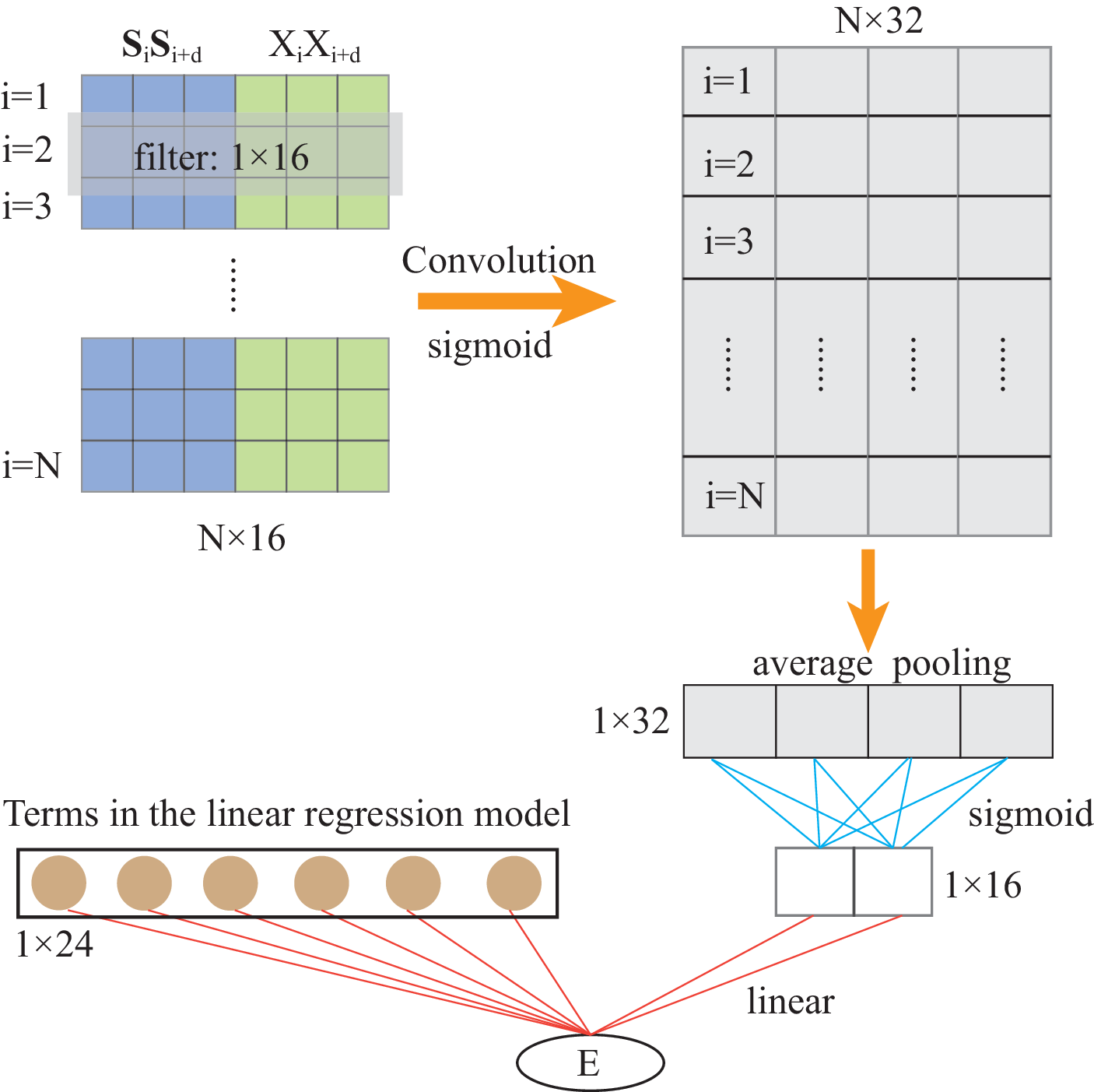}
\caption{ The architecture of the neural network model for predicting the free energy.
}
\label{Fig:fig5}
\end{figure}

\begin{figure}[t]
\centering
\includegraphics[width=0.99\columnwidth]{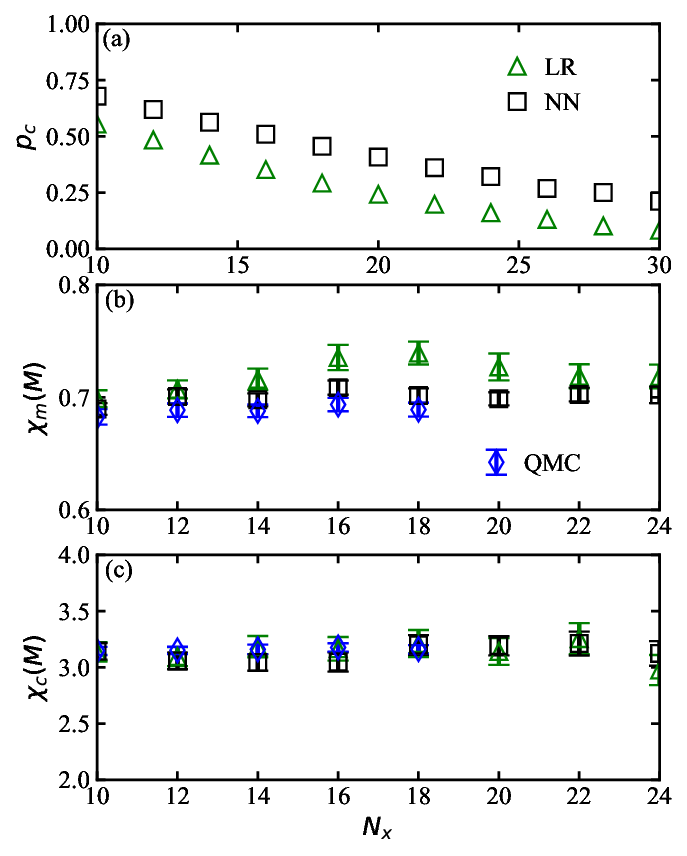}
\caption{(a) The cumulative update probability $p_c$ for the linear regression (LR) and neural network (NN) models. (b) The magnetic susceptibility $\chi_m(M)$ at the momentum point $M$ for different lattice sizes $N_x$. (c) The charge susceptibility $\chi_c(M)$ at the momentum point $M$ for different lattice sizes $N_x$. These results were obtained from a model with $g=1.5$. The temperature is set as $\beta=10$.
}
\label{Fig:fig6}
\end{figure}

\subsection{Neural network model}
To improve the sampling efficiency, we develop a neural network model to predict the free energy. The architecture of the neural network is shown in Fig.~\ref{Fig:fig5}. This network has two inputs. The first input consists of two-point correlation functions, specifically ${\bf S}_{{\bf r}_i}{\bf S}_{{\bf r}_i+{\bf d}}$ and $x_{{\bf r}_i}x_{{\bf r}_i+{\bf d}}$, defined across the entire lattice. In our simulations, we set $\bf d$ to be $\pm {\bf x}$, $\pm {\bf y}$, ${\bf x} \pm {\bf y}$, and $-{\bf x} \pm {\bf y}$. Thus, the first input is represented as a two-dimensional matrix with a size of $N \times 16$. We apply a convolutional filter to this input with a kernel size of $1 \times 16$, resulting in $N$ outputs for each channel. In our simulations, the number of channels in the first hidden layer is set to 32, yielding a matrix of size $N \times 32$. Next, we use an average pooling layer to reduce the dimensionality to $1 \times 32$. This is followed by a fully connected layer with a sigmoid activation function, producing an output of size 16. The second input consists of the terms from the linear regression model defined in Eq.~(\ref{eq: LRfunc}), comprising 24 elements. We concatenate this second input with the output of the fully connected layer and feed the combined result into the final neural network layer with a linear activation function. In this neural network, the first input includes two-point correlations. However, due to the nonlinearity of the sigmoid activation function, the linear relationship between the predicted free energy and two-point correlations is not guaranteed in the neural network. To preserve the linear component, we include two-point correlations as part of the second input features. Additionally, to constrain the atomic displacements, we include $W(x)$ in the second input. The significance of these second-input features is discussed in App.~\ref{sec:app3}.

The neural network model enhances training accuracy to some extent (see details in App.~\ref{sec:app3}). For example, at $g=1.5$ and $N=10$, the MSE loss function of the neural network model is approximately 0.014, which is smaller than the MSE loss function of the linear regression model (0.038). We incorporate this neural network model into the QMC method to simulate the Holstein-Spin-Fermion (HSF) model for various lattice sizes. Figure~\ref{Fig:fig6} illustrates the cumulative update probability $p_c$ for the SLQMC method using linear regression and neural network models. The results show that $p_c$ for the neural network model is larger than that of the linear regression model, indicating that the neural network model effectively improves sampling efficiency.

We compare the magnetic susceptibility and the charge susceptibility at $g=1.5$ from the SLQMC method using both the LR and NN models with the results from the standard QMC, as shown in Fig.~\ref{Fig:fig6}(b) and Fig.~\ref{Fig:fig6}(c). The SLQMC results from the LR and NN models are generally consistent, although $\chi_m(M)$ computed using the LR model is slightly larger than that from the neural network model. This minor discrepancy arises from differences in the predictive accuracy of these two models. When compared with the standard QMC results, $\chi_m(M)$ from the neural network model shows better agreement, indicating that the neural network model provides more accurate predictions than the LR model.

\subsection{Limitations of the machine learning model}
\begin{table}
\centering
\includegraphics[width=0.95\columnwidth]{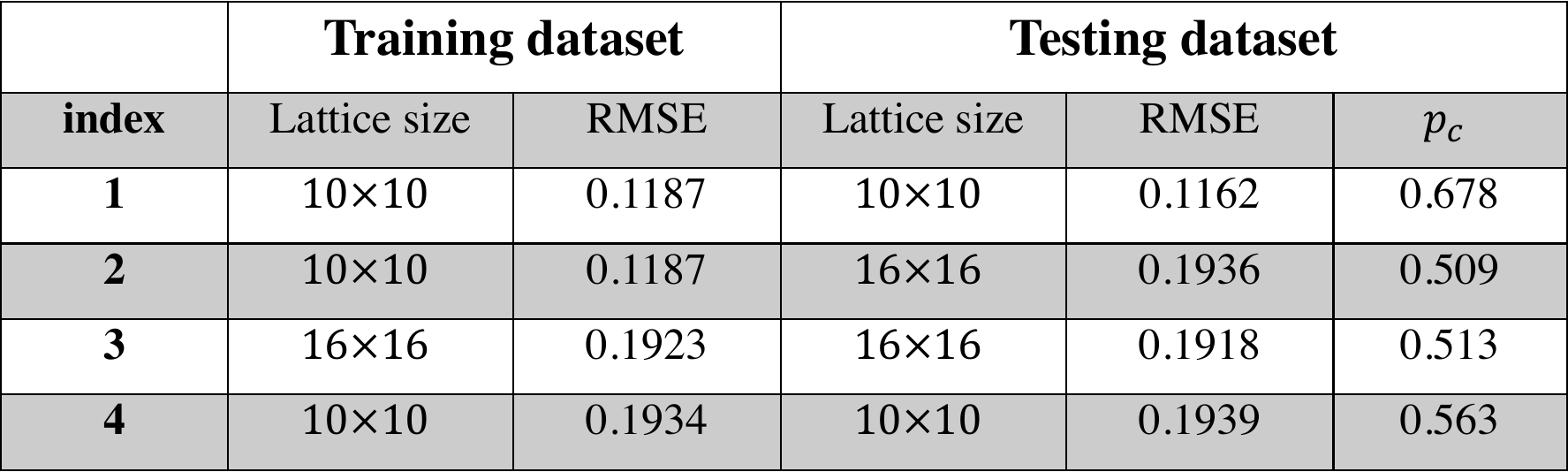}
\caption{A comparison of the root-mean-squared-error (RMSE) for different training and testing datasets at $g=1.5$ and $\beta=10$.}
\label{table:tab1}
\end{table}

\begin{figure}[t]
\centering
\includegraphics[width=0.95\columnwidth]{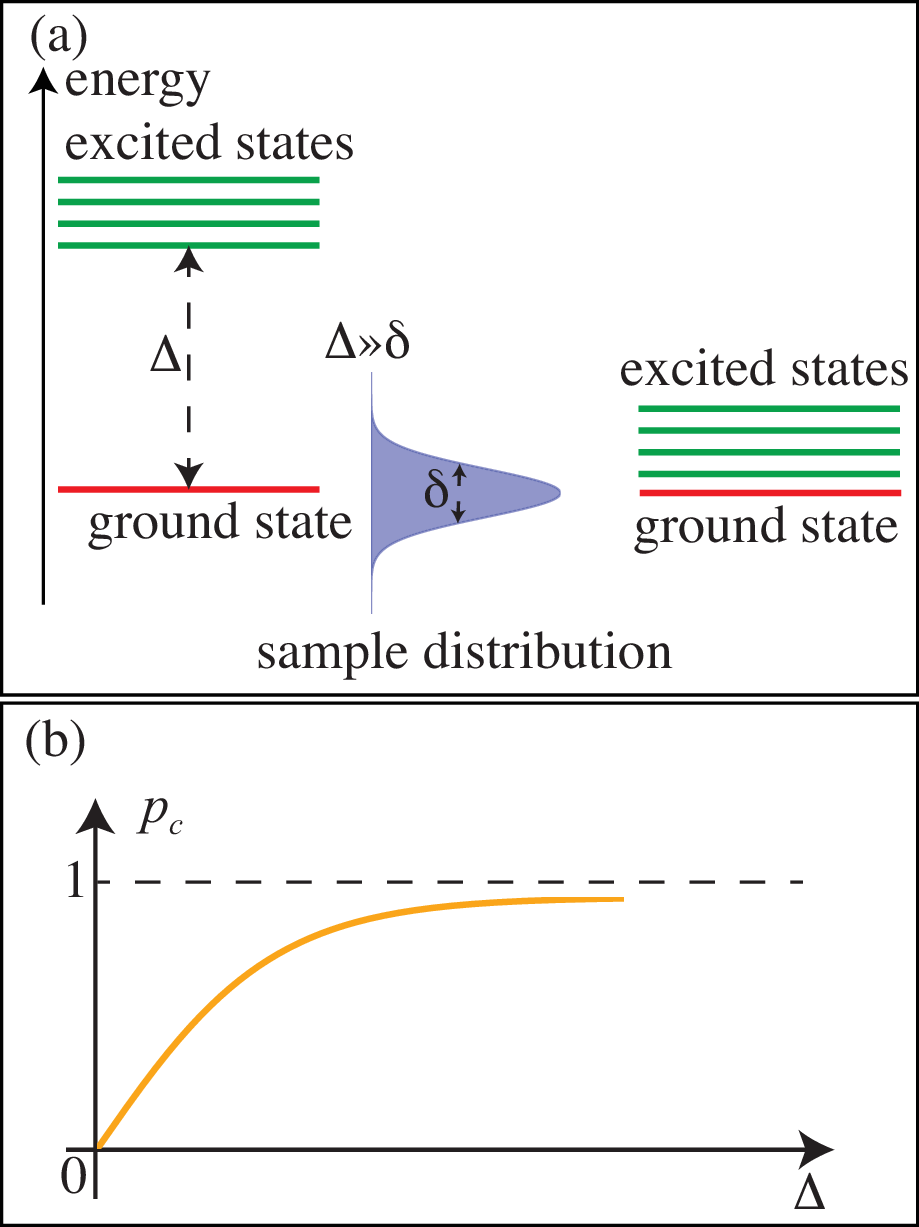}
\caption{(a) A sketch of the ground state and excited states. The Gaussian wave packet shows the distribution of quantum states generated from samples in the energy space. (b) A qualitative trend of the cumulative update probability $p_c$ and the energy gap between the ground and first excited states.
}
\label{Fig:fig8}
\end{figure}

Even though the training accuracy is increased by adopting the neural network model, the cumulative update probability $p_c$ remains suppressed by the lattice size. This suppression arises from two factors. The first is that RMSE increases as the network is applied to a large system. The second is the finite-size effect, wherein the energy gap between the ground state and low-energy excited states decreases as the lattice grows.

To illustrate the suppression of $p_c$, we analyze the RMSE across different training and testing datasets, as shown in Table~\ref{table:tab1}. We first trained a neural network using data from the 10×10 lattice, which yields a training RMSE of 0.1187. When tested on the 10$\times$10 and 16$\times$16 lattices, the RMSEs are 0.1162 and 0.1936, respectively. This result show that RMSE increases as the lattice size increases, leading to a reduction of the cumulative update probability from 0.678 to 0.509.  Next, we trained another neural network using samples from the 16$\times$16 lattice, which achieved a testing RMSE of 0.1918. The resulting $p_c$ of the second neural network is comparable to that of the first neural network, as both models exhibit similar RMSEs on the 16$\times$16 lattice. These results imply that the cumulative update probability depends on RMSE.

The finite-size effect also influences  $p_c$. To illustrate this, we trained a third network using data from the 10$\times$10 lattice, which achieved a testing RMSE of 0.193 on the same lattice. The cumulative update probability for this network for the 10$\times$10 lattice is 0.563, which is significantly larger than the value of $p_c$ ($\sim$0.51) of the other two networks applied to the 16$\times$16 lattice, despite their similar RMSEs. This discrepancy arises from the finite-size effect: as the lattice size increases, the energy gap $\Delta$ between the ground state and low-energy excited states decreases~\cite{Ueda2008Finite,Li2016Nonlocal}. When $\Delta$ is much larger than the RMSE of the machine learning model, the quantum state derived from the generated samples closely approximates the ground state, as shown in Fig.~\ref{Fig:fig8} (a). However, if the energy gap becomes smaller than the RMSE, the quantum state derived from the generated samples becomes a superposition of the ground and excited states. As the energy gap narrows, the probability of the excited states in this quantum state increases, leading to a decrease in the cumulative update probability $p_c$, as shown in Fig.~\ref{Fig:fig8} (b).  This framework explains the smaller $p_c$ values observed for the first two neural networks applied to the 16$\times$16 lattice. Therefore, an efficient sampling of SLQMC for a large lattice requires highly accurate machine learning models. Specifically, the RMSE loss function should be smaller than $1/N^{\alpha}$, where $\alpha$ depends on the details of the system~\cite{Blote1986Donformal}.

\section{Temperature dependence}
\begin{figure}[t]
\centering
\includegraphics[width=0.99\columnwidth]{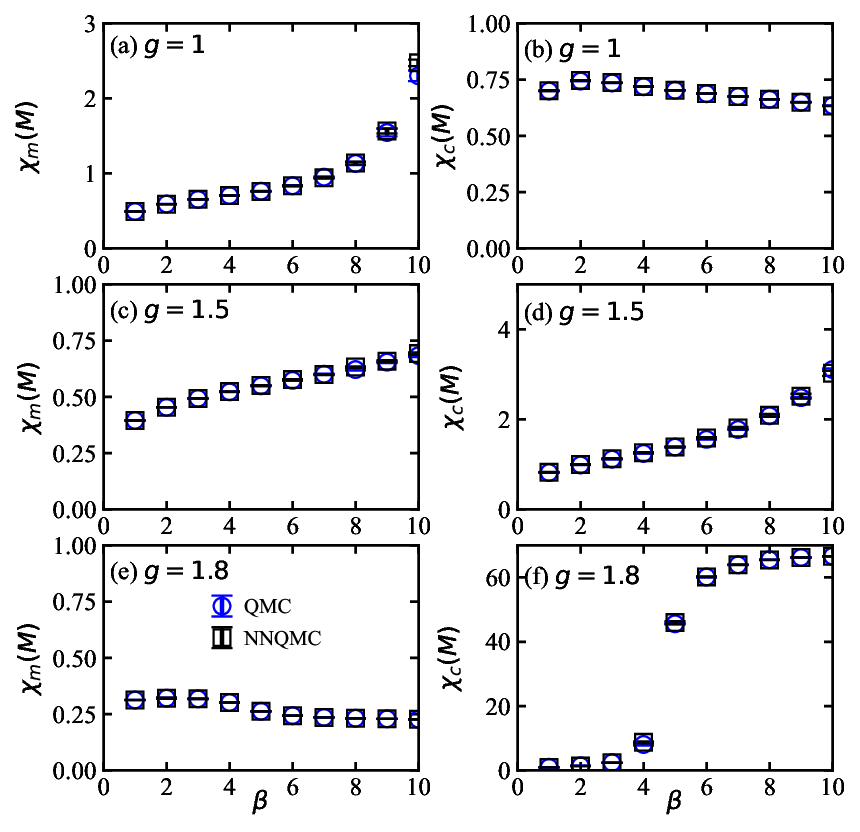}
\caption{The left three panels show the temperature-dependent magnetic susceptibility $\chi_m$ at the momentum point $M$ for $g=1.0$, $g=1.5$, and $g=1.8$, respectively. The right three panels plot the temperature-dependent charge susceptibility $\chi_c$ at the momentum point $M$. The blue and red symbols denote the results of the standard QMC and the neural network QMC.
}
\label{Fig:fig9}
\end{figure}

\begin{figure}
\centering
\includegraphics[width=0.99\columnwidth]{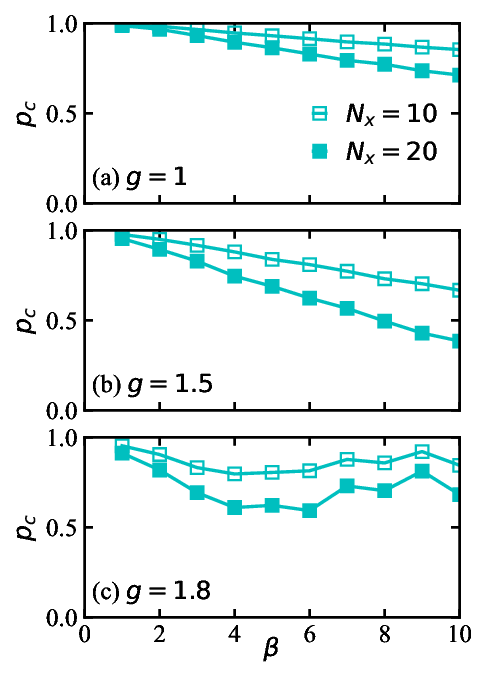}
\caption{(a), (b), and (c) plot the cumulative update probability $p_c$ at different temperatures for $g=1.0$, $g=1.5$, and $g=1.8$, respectively. The neural network is trained using samples generated from a $10\times10$ lattice. 
}
\label{Fig:fig10}
\end{figure}

Finally, we evaluate the performance of the neural network-assisted QMC method across different temperatures. To achieve this, we collected samples from QMC simulations conducted at various temperatures and trained the neural network at each temperature. Figure~\ref{Fig:fig9} compares the results of the standard QMC method with those of the neural network-assisted QMC for $g=1$, $g=1.5$, and $g=1.8$. The three panels on the left depict the temperature-dependent magnetic susceptibility at the $M$ point, while the three panels on the right illustrate the temperature-dependent charge susceptibility at the same momentum point. The close agreement between the neural network-assisted QMC and the standard QMC results demonstrates that the neural network-assisted method is reliable and effective in both the low- and high-temperature regimes.

We further analyze the temperature-dependent cumulative update probability at $g=1.0$, $g=1.5$, and $g=1.8$, shown in Fig.~\ref{Fig:fig10}. For $g=1.0$ and $g=1.5$, $p_c$ decreases with decreasing temperature, which can be attributed to the increase in the RMSE (see details in App.~\ref{sec:app2}). Notably, $p_c$ at $g=1.5$ decreases more rapidly than that at $g=1.0$, as the RMSE is larger at low temperatures in the former case. In contrast, $p_c$ at $g=1.8$ displays a non-monotonic trend. $p_c$ initially decreases as $\beta$ increases from 1 to 4 and then increases as $\beta$ increases further. This non-monotonic behavior arises from the competition between the RMSE and $\beta$ because the cumulative update probability relies on $\beta E_F$. At $g=1.8$, the RMSE decreases monotonically from 0.0307 to 0.0014 as $\beta$ increases from 1 to 10. However, the product $\beta\times\text{RMSE}$ exhibits a non-monotonic behavior, aligning with the behavior of $p_c$ (see details in App.~\ref{sec:app2}.). By examining the behavior of the RMSE and $\beta\times\text{RMSE}$ at lower temperatures and other values of $g$, we find that the non-monotonic behavior of $\beta\times\text{RMSE}$ arises from the temperature-induced phase transition (see details in App.~\ref{sec:app4}). Slightly above the phase transition temperature, the system exhibits strong fluctuations. Below the phase transition temperature, the system becomes an ordered state. Compared with the strongly fluctuating state, machine learning achieves better performance in the ordered state, yielding smaller RMSE values and consequently a decrease in $\beta\times\text{RMSE}$.

\section{Conclusion}\label{sec:IV}
Our work develops a self-learning QMC method to handle the interplay between the {\eph} and spin-spin interactions. We examine the performance of this SLQMC by simulating the classical Holstein-spin-fermion model. The results of SLQMC using a linear regression model and the results of the standard QMC are found to be consistent across the phase transition from the AFM state to the CDW state. However, we observe that the sampling efficiency of SLQMC is suppressed near the phase transition point, which is attributed to the increased MSE of the linear regression model in this regime. Replacing the linear regression model with a neural network model reduces  the MSE and improves the sampling efficiency. Despite this improvement, the sampling efficiency of SLQMC is suppressed as the lattice size increases. Our analysis reveals that this suppression originates from the increased RMSE when the network is applied to a large system and from the shrinking energy gap between the ground state and low-energy states decreases as the lattice grows.

Our self-learning QMC method is well-suited for large-scale simulations of quantum systems. For instance, it can enable large-scale studies of stripe order in a three-orbital spin-fermion model for cuprates~\cite{Hussein2019Half}, and extends the research to a three-orbital Holstein-spin-fermion model to explore phonon effects on stripe formation~\cite{Karakuzu2020stripe}. Furthermore, this method can be used to study three-dimensional quantum systems, such as investigating charge order in BaBiO$_3$~\cite{Jiang2021Polaron} and examining how electron-electron interaction influence bipolaron formation. 
While the hybrid quantum Monte Carlo method can also simulate three-dimensional system~\cite{Cohen2023Ahybrid}, however, it has difficulties in handling electron-electron or spin-spin interactions. Moreover, self-learning QMC method can be applied to obtain spectroscopies with high momentum resolution, which is essential for comparison with experimental data~\cite{LeBlanc2019Magnetic, Li2020Magnetic}.

\begin{figure}
\centering
\includegraphics[width=0.99\columnwidth]{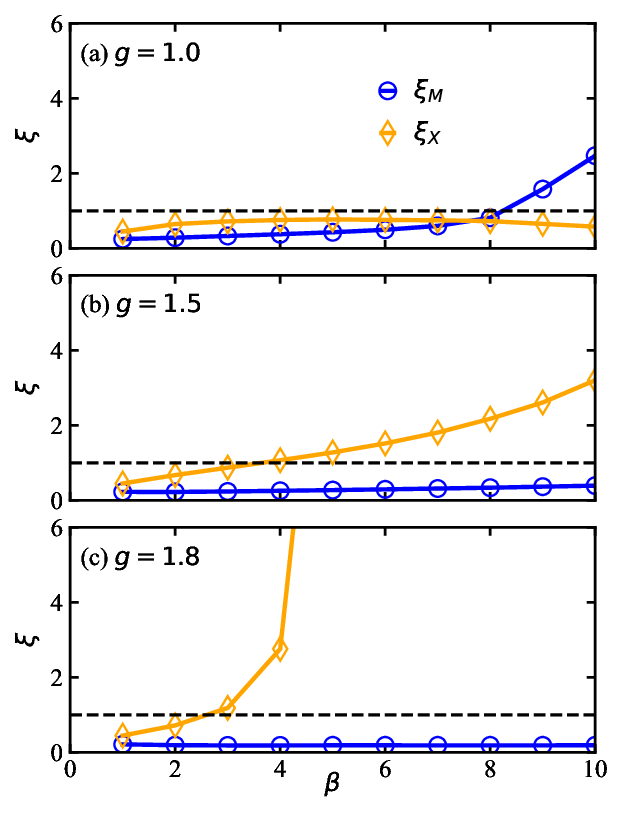}
\caption{ Coherence length for the spin correlation function and the displacement correlation function. $\xi_M$ and $\xi_X$ denote the coherence length for the spin and displacement, respectively. Results are obtained from the QMC simulations on a $10\times10$ lattice with $J=2$ and $K=1$.
}
\label{Fig:appd1}
\end{figure}

\begin{figure*}[t]
\centering
\includegraphics[width=0.9\textwidth]{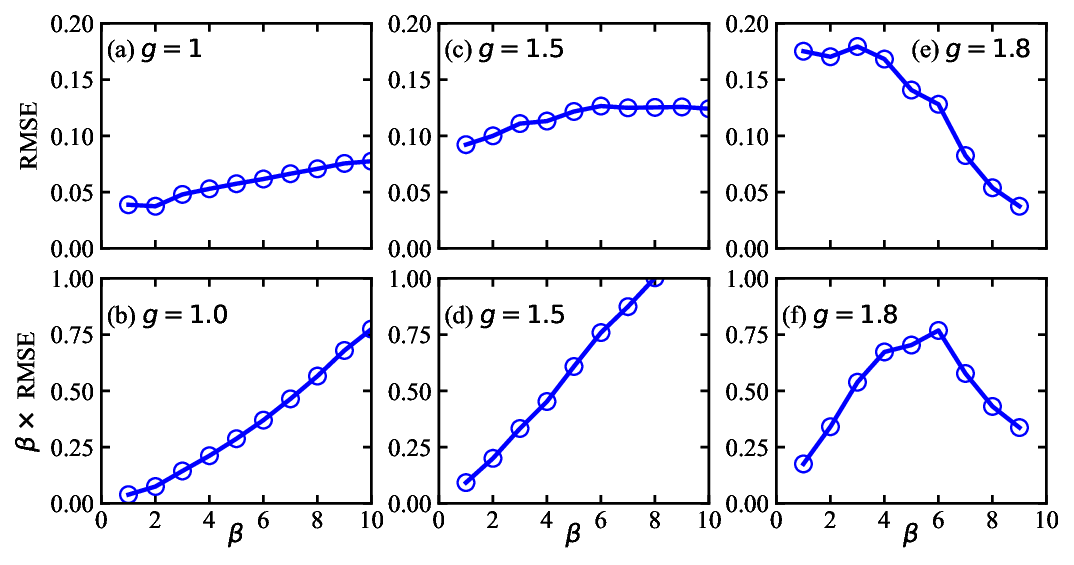}
\caption{ The top panel shows the root-mean-squared-error (RMSE) of the neural network trained on the samples for $g=1.0$, $g=1.5$, and $g=1.8$ at different temperatures. The bottom panel shows the corresponding values of $\beta\times\text{RMSE}$ as a function of $\beta$.}
\label{Fig:appd2}
\end{figure*}

\section*{Acknowledgement} 
We acknowledge funding support from the  National Natural Science Foundation of the People's Republic of China (Grand No. 12204236). The authors also acknowledge the startup funding support from Northeastern University, Shenyang.

\appendix
\section{Coherence Length}\label{sec:app1}

\begin{figure}[t]
\centering
\includegraphics[width=0.9\columnwidth]{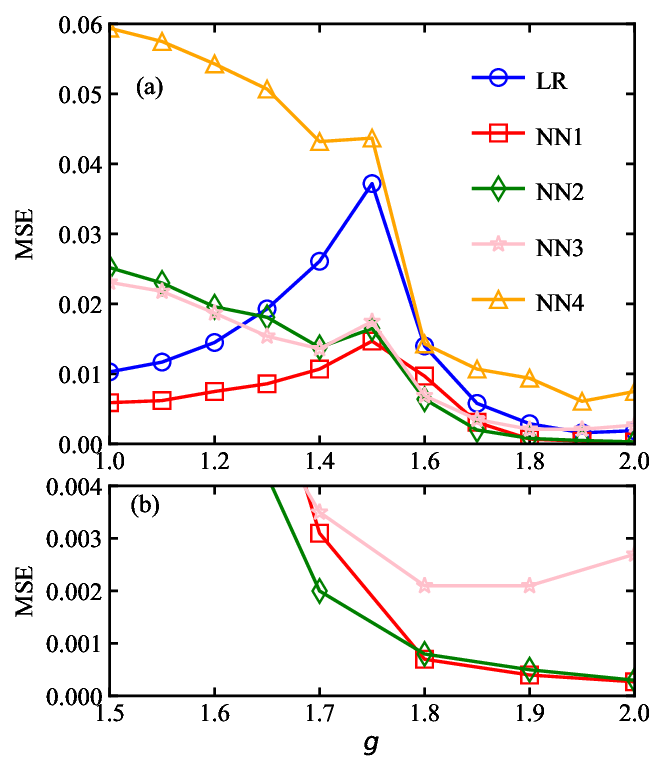}
\caption{ A comparison of the mean-squared-error (MSE) of five different machine learning models. Here, LR stands for the linear regression model. NN1 denotes the neural network adopted in the main text. NN2, NN3, and NN4 share the same architecture as NN1, but differ in their second inputs. The second input of NN2 includes only $W(x)$. The second input of NN3 includes only $\sum_i x_{{\bf r}_i}^2$. NN4 has no second input.
}
\label{Fig:appd3}
\end{figure}

\begin{figure}[t]
\centering
\includegraphics[width=0.8\columnwidth]{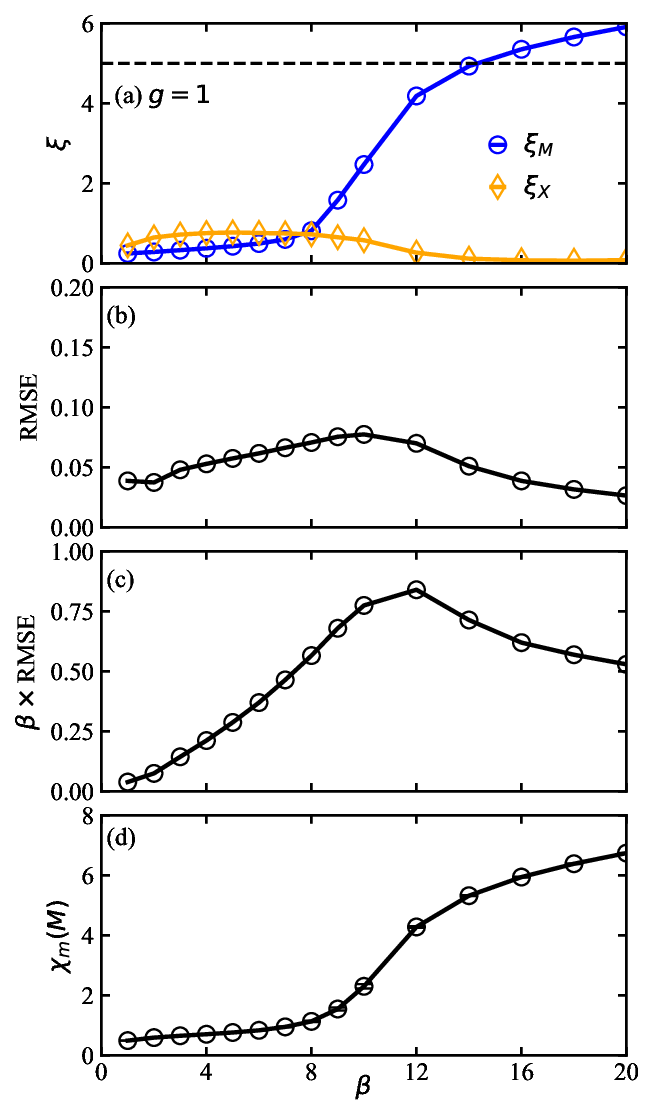}
\caption{(a) Coherence length for the magnetic ($\xi_M$) and displacement ($\xi_X$) correlation functions at $g=1.0$. (b) The mean-squared-root-error (RMSE) as a function of temperature at $g=1.0$. (c) The corresponding $\beta\times$RMSE as a function of temperature. (d) The magnetic susceptibility $\chi_m(M)$ at the momentum M point as a function of temperature.
}
\label{Fig:appd4}
\end{figure}

\begin{figure}[t]
\centering
\includegraphics[width=0.8\columnwidth]{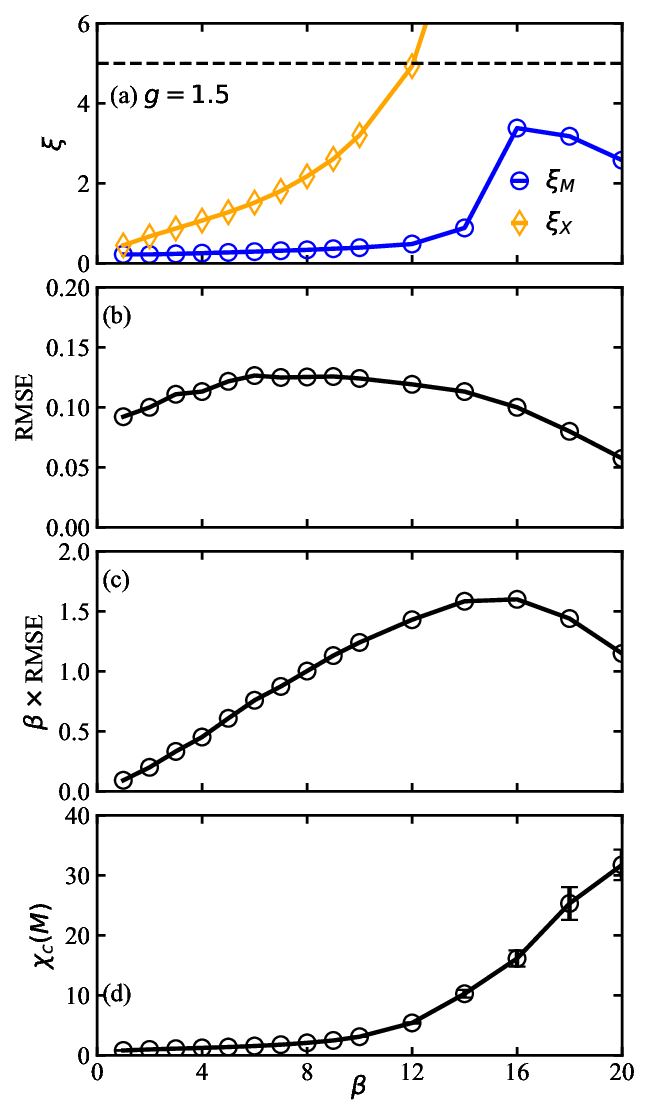}
\caption{(a) Coherence length for the magnetic ($\xi_M$) and displacement ($\xi_X$) correlation functions at $g=1.5$. (b) The mean-squared-root-error (RMSE) as a function of temperature at $g=1.5$. (c) The corresponding $\beta\times$RMSE as a function of temperature. (d) The charge susceptibility $\chi_c(M)$ at the momentum M point as a function of temperature.
}
\label{Fig:appd5}
\end{figure}

We analyze the coherence lengths of the spin correlation function and the displacement correlation function across different temperatures and $\eph$ interaction strengths. Both functions exhibit a momentum dependence that can be described by a Lorentzian form:
\begin{eqnarray}
\chi({\bf q})=\frac{A}{(q_x-Q_x)^2+(q_y-Q_y)^2+1/\xi^2},
\end{eqnarray}
where $(Q_x, Q_y)$ is the ordering wavevector, and $\xi$ represents the coherence length. In our analysis, the ordering wavevector for both the AFM state and the CDW state is $(\pi/a, \pi/a)$. Here, we denote the coherence lengths for the spin and displacement correlation functions as $\xi_M$ and $\xi_X$, respectively.

Figure~\ref{Fig:appd1} presents the behavior of $\xi_M$ and $\xi_X$ across various temperatures and $\eph$ interaction strengths, as determined from QMC simulations conducted on a $10\times10$ lattice with $J=2$ and $K=1$. At $g=1.0$, $\xi_X$ remains below 1 (indicated by the black dashed line), suggesting a disordered displacement configuration. In contrast, $\xi_M$ gradually increases as the temperature decreases, indicating the appearance of AFM fluctuations. However, at $\beta=10$, $\xi_M$ remains below 5 (the half-width of the lattice), indicating that an ordered state does not form at this temperature.

At $g=1.5$, $\xi_M$ remains below 1 throughout the range from $\beta=1$ to $\beta=10$, indicating a disordered spin configuration. In contrast, $\xi_X$ increases smoothly with $\beta$, signaling the development of checkerboard displacement fluctuations. However, at $\beta=10$, $\xi_X$ remains below 5, demonstrating that an ordered checkerboard state does not form. At $g=1.8$, the spin configuration remains disordered, as evidenced by $\xi_M$ being less than 1. In contrast, $\xi_X$ increases rapidly with decreasing temperature, surpassing 5 near $\beta=4.5$, which indicates the formation of the checkerboard order.

\section{Training results of the neural network}\label{sec:app2}
We present the RMSE of the trained neural network for $g=1.0$, $g=1.5$, and $g=1.8$ at various  temperatures in the top panel of Fig.~\ref{Fig:appd2}. For $g=1.0$ and $g=1.5$, the RMSE generally increases with $\beta$. In contrast, for $g=1.8$, the RMSE trends to  decrease as $\beta$ increases. The bottom panel of Fig.~\ref{Fig:appd2} shows the corresponding values of $\beta\times\text{RMSE}$. For $g=1.0$ and $g=1.5$, $\beta\times\text{RMSE}$ increases as the temperature decreases, leading to a suppression of the accumulated probability in the self-learning QMC. In the case of $g=1.8$, $\beta\times\text{RMSE}$ exhibits a non-monotonic trend: the value increases as $\beta$ rises from 1 to 4, then decreases as $\beta$ increases further. This non-monotonic behavior aligns with the trend of $p_c$ observed in Fig.~\ref{Fig:fig10}(c).

\section{A comparison of different machine learning models}\label{sec:app3}
In the main text, we adopt two machine learning models: a linear regression model (LR) and a neural network model (NN1). Here, we investigate three additional neural network models — NN2, NN3, and NN4 — that share the same architecture as NN1, but differ in their second inputs. Specifically, the second input of NN2 includes only $W(x)$, that of NN3 includes only $\sum_i x_{{\bf r}_i}^2$, and NN4 has no second input. Figure~\ref{Fig:appd3} presents the MSE of all five models as a function of the electron–phonon coupling strength $g$. Among them, NN1 achieves the best performance, outperforming the LR model across all $g$. The MSE of NN2 is larger than that of NN1, indicating that the free energy has an important linear dependence on the two-point correlations.  In the AMF-fluctuating state, the MSE of NN2 is larger than that of NN3, whereas in the CDW state, NN2 outperforms NN3, suggesting that a complex effective onsite elastic energy is needed in the CDW phase. To maintain a uniform neural network architecture across all $g$, we adopt $W(x)$ rather than $\sum_i x_{\bf{r}_i}^2$ in the main text, sicne $W(x)$ captures the double-well lattice potential characteristic of a CDW state. NN4 consistently performs the worst, with its MSE higher than that of all other model, implying the necessity of  incorporating effective onsite elastic energy for accurate free energy predictions.

\section{Phase transition at $g=1.0$}\label{sec:app4}
Here, we present the magnetic phase transition at $g=1.0$ on a $10\times10$ lattice. Figure~\ref{Fig:appd4}(a) shows the coherence lengths of the magnetic ($\xi_M$) and displacement ($\xi_X$) correlations. It is found that $\xi_M$ increases continuously with decreasing temperature and exceeds 5 once $\beta>14$, implying the onset of magnetic order around $\beta=14$. Figures~\ref{Fig:appd4}(b) and~\ref{Fig:appd4}(c) plot the RMSE and $\beta\times$RMSE as functions of temperature. Both quantities increase with $\beta$ at first but begin to decrease once $\beta>14$. This behavior suggests that the machine learning model achieves better performance in the ordered state than in the strongly fluctuating regime. The same conclusion can also be obtained from results at $g=1.1$, $g=1.2$, $g=1.3$, and $g=1.4$.

Figure~\ref{Fig:appd4}(d) plots the magnetic susceptibility $\chi_m(M)$ at the momentum point M as a function of temperature. $\chi_m(M)$ increases continuously with decreasing temperature.

\section{Phase transition at $g=1.5$}\label{sec:app5}
We present the charge order phase transition at $g=1.5$ on a $10\times10$ lattice in Fig.~\ref{Fig:appd5}. Figure~\ref{Fig:appd5}(a) shows the coherence lengths of the magnetic ($\xi_M$) and displacement ($\xi_X$) correlations. It is found that $\xi_X$ increases continuously as temperature decreases and exceeds 5 once $\beta>12$, implying the onset of charge order around $\beta=12$. Interestingly, we observe that spin fluctuations are enhanced at $\beta=14$ and $\beta=16$ despite the presence of charge order. These fluctuations are eventually suppressed at lower temperatures. The enhanced spin fluctuation is attributed to the mean-field treatment of the electron-electron interaction. Figures~\ref{Fig:appd5}(b) and~\ref{Fig:appd5}(c) display the RMSE and $\beta\times$RMSE as functions of temperature. Both quantities increase continuously with decreasing temperature for $\beta<12$. The RMSE begins to decrease at $\beta=12$, consistent with the critical temperature of the phase transition, while the decrease of $\beta\times$RMSE sets in at $\beta=16$.

In the fluctuating state at $\beta=10$, the RMSE at $g=1.5$ is larger than that at $g=1.0$, resulting in a smaller value of $p_c$ at $g=1.5$. In the ordered state at $\beta=20$, the RMSE at $g=1.5$ remains larger than that at $g=1.0$. The corresponding $p_c$ values are 0.932 for $g=1.0$ and 0.633 for $g=1.5$. The larger RMSE at $g=1.5$ is attributed to the presence of strong spin fluctuations.

Figure~\ref{Fig:appd5}(d) plots the charge susceptibility $\chi_c(M)$ at the momentum point M as a function of temperature. $\chi_c(M)$ increases continuously with decreasing temperature.

\bibliography{main}
\end{document}